\documentclass[11pt]{article}
\usepackage{moriond,epsfig}
\usepackage{amsfonts, amsmath, amssymb}

\newcommand{\beqs}{\begin{eqnarray}}
\newcommand{\eeqs}{\end{eqnarray}}
\newcommand{\lb}[0] { \left( }
\newcommand{\rb}[0] { \right) }
\newcommand{\citesp}[1] {\hspace{0.8mm}\cite{#1}}

\begin{document}

\vspace*{4cm}
\title{NEUTRINO PRODUCTION IN NUCLEONIC INTERACTIONS IN GAMMA-RAY BURSTERS}

\author{HYLKE B. J. KOERS}

\address{Service de Physique Th{\'e}orique, Universit{\'e} Libre de Bruxelles (U.L.B.), CP225, Bld. du Triomphe, B-1050 Bruxelles, Belgium}

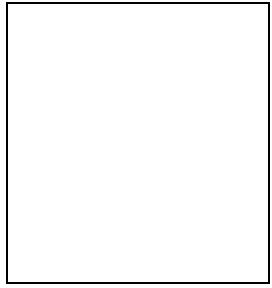
\maketitle\abstracts{Neutrinos produced in gamma-ray bursters (GRBers) may provide a unique probe for
the physics of  these extreme astrophysical systems. Here we discuss neutrino production in \mbox{inelastic}
neutron-proton collisions within the relativistic outflows associated with GRBers. We consider both the widely
used fireball model and a recently proposed magneto-hydrodynamic (MHD) model for the
GRB outflow.}

\section{Introduction}

Gamma-ray bursts (GRBs) are short and energetic flashes of
gamma rays ($\sim100$ keV), reaching Earth from apparently random direction
at a rate of a few per day (see Ref.\citesp{Meszaros:2006rc} for a review). The luminosity of these bursts may be
very large, sufficient to temporarily outshine
all other gamma-ray sources combined. 
Following their accidental
discovery in 1967, the origin of these remarkable events has puzzled astronomers
for three decades. In particular, the question whether gamma-ray bursts were
produced by sources within our galaxy or at cosmological
distances has remained under debate until the 1990s.

In the last decade there has been tremendous
progress in our understanding of GRBs. This is largely due
to observations of GRB afterglows -- periods of prolonged broad-band electromagnetic
emission following the actual burst -- that were first discovered in 1997. 
The cosmological distance scale has been established by redshift measurements
of the afterglow, and in some cases afterglow
observations have allowed for identification of the host galaxy, providing
further clues as to the nature of gamma-ray bursters (GBRers).  There is presently compelling evidence that long-duration GRBs
(the subclass of GRBs lasting more than 2 sec) are ultimately caused
by core-collapse of massive stars\citesp{Woosley:2006fn}, although the situation
for short-duration GRBs (lasting less than 2 sec) is less clear.

One can easily estimate that the energy released in a stellar core-collapse matches
that required to power a GRB, but the mechanism responsible for the energy transfer is far from obvious.
The widely accepted framework to describe this is divided into four phases.
In the \emph{initial phase}, core-collapse of the massive star results
in a black hole-accretion disk system. This launches a jet, a collimated
outflow of plasma that contains a small
baryonic component. In the \emph{accelerating phase}, this plasma 
accelerates to a velocity close to that of light (Lorentz factor $\sim$ 300).
In this acceleration process, the initial energy of the plasma is transfered
to bulk kinetic energy of the baryons that are contained in the plasma.
In the \emph{coasting phase}, the outflow moves with a fixed velocity through the pre-burst stellar environment.
Here dissipation of the kinetic energy in the flow, most likely as synchrotron
emission of shock-accelerated electrons, gives rise to the actual GRB.
Finally, the afterglow is attributed to the interaction of the outflow
with the external medium during the \emph{afterglow phase}.

Although the framework described above successfully explains the general features of the  observations,
many questions remain. Arguably one of the most important issues is the nature of the  
relativistic outflow. Within the widely used fireball model, it is understood that the plasma is initially
dominated by thermal energy. Alternatively, the
energy may be predominantly in electromagnetic form. Such outflows
are expected naturally when a magnetized accretion disk is surrounding
the central black hole.\citesp{Thompson:1994zh,Meszaros:1996ww,Spruit:2000zm}
Further questions concern for example the initial collimation of the flow, where magnetic fields may also
play an important dynamical role,
and the details of the energy dissipation process, which is likely to involve some
particle acceleration mechanism such as shock acceleration.

Besides the intrinsic motivation to better understand the physics of GRBers, further 
motivation is provided by the connection to other fields of physics.
Since GRBers are believed to be efficient astrophysical particle accelerators,
they are candidate sources of high-energy neutrinos and cosmic rays
and provide a laboratory to study the acceleration mechanism.
Furthermore, it has been proposed to use GRBs as standard candles to constrain the
evolution of the universe.\citesp{Ghirlanda:2006ax}  Finally, there are more speculative proposals, e.g.
to use the arrival times of low- and high-energy emission to constrain Lorentz
violating interactions.\citesp{AmelinoCamelia:1997gz}
 
Neutrinos are promising probes of the environment of GRB sources.
Neutrino emission is complementary to the electromagnetic
emission in two respects. First, neutrinos mostly trace the hadronic component
of GRB outflows whereas electromagnetic radiation mostly traces the
leptonic component. Second, neutrinos can leave the GRB source
when it is still optically thick.
Substantial neutrino production may be expected in various phases of a developing GRB.
In the initial phase, neutrino emission can constrain the formation of 
GRB fireballs. Within the fireball model, the dominant neutrino production
process in this phase is electron-positron annihilation (providing a counterexample to the mostly hadronic
production mechanisms). Under favorable circumstances,
this may give rise to copious neutrino production. However, this mechanism
is not sufficiently efficient to carry away the bulk of the
fireball energy or to qualitatively modify the dynamical behavior of the fireball.\citesp{Koers:2005ya}
In both the coasting and afterglow phases of a developing GRB, it is believed that 
kinetic energy is dissipated through shock acceleration of electrons. These shocks
will likely also accelerate any protons contained in the fireball. Interactions of
these high-energy protons with target nucleons or photons give rise to a flux
of high-energy
neutrinos that offers good detection prospects with the upcoming km$^3$ neutrino
detectors such as IceCube.\citesp{Waxman:1997ti,Waxman:1999ai} These neutrinos provide information on the nature of the flow, in particular
the strength of the hadronic component, and on the energy dissipation process.

Here we report on a different mechanism to create neutrinos in GRBers, namely 
inelastic neutron-proton ($np$) collisions that occur during the accelerating phase. 
We compare the typical neutrino emission through this mechanism
for two competing models: the fireball model and the recently proposed `AC' model\citesp{Spruit:2000zm,Giannios:2004di},
which assumes that the energy in the outflow is predominantly electromagnetic.
The motivation of this work is to estimate the
detection prospects of this neutrino emission and to investigate whether it can be used to differentiate between
the fireball model and the AC model.
The $np$ mechanism has been considered within the fireball model \mbox{before.\citesp{1999ApJ...521..640D,Bahcall:2000sa}} Our estimates are more pessimistic than existing ones, which can be traced
to the more accurate modeling of the inelastic $np$ cross section adopted in our work. For the AC model, the
mechanism was first considered in Ref.\citesp{Koers:2007ww}, which forms the basis of the present discussion.

In the following section we discuss the dynamics of GRB outflows containing neutrons and protons,
both within the fireball model and within the AC model. We then discuss neutrino production
through $np$ interactions, and finally we present our conclusions.

\section{Dynamics}
\subsection{Acceleration in the fireball model and the AC model}
A striking feature of GRB models is the bulk relativistic motion.
This ingredient is motivated by an observational paradox: the short timescales and large energies 
suggest a huge energy density and thus an optically thick source. This then implies 
that the photon spectrum should be quasi-thermal, while observations show that it is not.
Relativistic motion solves this problem by increasing the physical timescale compared to
that inferred from observations, and by decreasing the photon energy in the source compared
to the observed energy. The mechanism
to accelerate the flow to relativistic velocities differs between models. In the fireball model, acceleration
results from the pressure that photons exert on the optically thick fireball.  In this case
the dynamics of the flow may be approximated with\citesp{Paczynski:1986px}
\beqs
\label{eq:GR:FB}
\Gamma \propto r \, ,
\eeqs
where $\Gamma$ is the Lorentz factor of the flow, and $r$ the radius of the flow (i.e., the distance
from the central black hole). 
In the AC model, the energy to accelerate the outflow is provided by magnetic reconnection,
a mechanism that converts electromagnetic energy into heat and bulk motion.
When the magnetic field lines predominantly change polarity in the flow direction, as we will assume,
the dynamics of the flow may be approximated with\citesp{2002A&A...387..714D}
\beqs
\label{eq:GR:AC}
\Gamma \propto r ^{1/3}\, .
\eeqs
Comparison with eq. \eqref{eq:GR:FB} shows that the acceleration of the flow is much more
gradual in the AC model than in the fireball model. As we will see, this directly affects the
neutrino flux from $np$ collisions.

In both the fireball model and the AC model, acceleration of the flow stops when there is no more 
energy available to further accelerate the baryons. In the fireball model, the acceleration of the flow
can also be terminated when the flow, whose energy density decreases with increasing radius, 
becomes optically thin. 

\subsection{Neutron-richness}
Since the baryons that are contained in the flow are to be accelerated to high Lorentz factors, 
the initial baryon density cannot be too large.
This requirement is generally stated in terms of  a dimensionless \emph{baryon loading
parameter}
\beqs
\label{eq:defeta}
\eta \equiv  L / \dot{M} c^2 \sim 10^3 \, ,
\eeqs
where $L$ denotes the total luminosity of the flow and $\dot{M}$ the mass flux.
Near the central black hole, the
typical energy density is larger than nuclear binding energies so that the baryonic component
will consist predominantly of free protons and neutrons. The ratio of neutrons to protons at the base
of the outflow
is determined by the competition of electron capture on protons and positron capture on neutrons.
Recent studies\citesp{Chen:2006rra}
favor a neutron-rich environment, so that we expect that the outflow associated with a
developing GRB is initially  also neutron-rich. 
The neutron-to-proton ratio is parameterized with
\beqs
\label{eq:defxi}
\xi \equiv  \dot{M}_n / \dot{M}_p \sim 1 \, ,
\eeqs
where $\dot{M}_{n (p)}$ denotes the neutron (proton) mass flux.
At larger radii, where the energy densities
are smaller, nucleosynthesizing reactions reduce the number of free neutrons. However,
a significant amount of neutrons is expected in the flow up to the radius where neutron
decay becomes important. This radius  is much larger than the radii relevant to $np$ collisions
and thus neutron decay is not important for the mechanism considered in this work.

\subsection{Neutron decoupling and pion production}
Eqs. \eqref{eq:GR:FB} and \eqref{eq:GR:AC} are idealized approximations
that are only valid when the baryons contained in the plasma play no dynamical role. 
Detailed numerical studies\citesp{Koers:2007ww,Rossi:2005uc} indicate that a reasonably strong baryonic component affects
the dynamics. However, eqs.  \eqref{eq:GR:FB} and \eqref{eq:GR:AC}  provide a reasonable approximation
to the full dynamical behavior that captures  the properties which are essential to the particle
production problem discussed here. We will thus  neglect the dynamical importance of
nucleons in this section.

Regardless of the mechanism that accelerates the flow, protons are strongly coupled
to the other plasma components by electromagnetic interactions and follow the dynamics
of the flow. The neutrons, on the other hand, are only coupled to the plasma through inelastic
$np$ collisions. The nucleon number densities are initially very large so that the $np$ interaction timescale
is much shorter than the dynamical timescale. In this regime, the neutrons and protons essentially behave
as a single fluid.
As the outflow expands, the number densities
decrease and the scattering timescale increases.
When the $np$ scattering timescale becomes smaller than
the dynamical timescale, the neutrons  decouple from the plasma and
coast with a certain terminal velocity.

When the flow is still in the accelerating phase at $np$ decoupling, the protons are accelerated
further and consequently a bulk velocity difference develops between protons and neutrons.
If this velocity becomes sufficiently large, pions can be created in inelastic $np$  collisions. The threshold
condition to produce pions may be expressed as
$\chi \equiv \Gamma_p / \Gamma_n > \chi_\pi \equiv 2.15 $ ,
where $\Gamma_{p (n)} $ denotes the proton (neutron) Lorentz factor. Approximating the dynamics
of the outflow with $\Gamma \propto r^p$ (where $p=1$ corresponds to the fireball model and
$p=1/3$ to the AC model), we observe that the radius where pion production occurs $r_\pi$ and
the decoupling radius $r_{np}$ are related through
$ r_\pi \simeq r_{np} \chi_\pi^{1/p} $.
Hence, in the fireball model the pion production radius is roughly twice the decoupling radius,
while in the AC model it is an order of magnitude larger.

If the outflow contains many baryons, the available amount of energy per baryon is relatively small. In this
case the acceleration of the flow may saturate before $np$ decoupling, thus preventing
inelastic collisions. Hence
a sufficiently `pure' flow ($\eta \gtrsim 500$ for the fireball model, or $\eta \gtrsim 200$ for the AC model)
is required for particle production in inelastic $np$ collisions.

\section{Particle production in neutron-proton collisions}

\subsection{Interaction probability}
The probability $d \tau$ for
a neutron moving with dimensionless velocity $\beta_n$ to interact
with a proton population moving with dimensionless velocity $\beta_p$,
within an infinitesimal radius $r \ldots r+dr$ is\citesp{Koers:2007ww}
\beqs
\label{eq:dtaudr}
d \tau = \sigma  \Gamma_{p} n'_{p}  \lb \frac{\beta_{p} - \beta_{n}}{\beta_{n}} \rb d r
\simeq  \frac{\sigma  n'_{p} }{2 \Gamma_{n}} \lb \chi - \frac{1}{\chi} \rb  d r \, ,
\eeqs
where  $n'_p$ denotes the comoving proton density, $\sigma$ is the inelastic $np$ cross
section \footnote{We refer the reader to Ref.\citesp{Koers:2007ww} for the adopted approximation for $\sigma$.},
and we have assumed that $\beta_{n} \simeq 1$ and $\beta_{p} \simeq 1$  in the second equality.
For outflows that follow an acceleration profile $\Gamma \propto r^p$ up to infinity, 
integrating eq. \eqref{eq:dtaudr} gives the probability $\tau$ for an inelastic $np$ collision to occur somewhere
between the pion production radius and infinity.
The result is independent of any model parameters except the index $p$. Performing this integral, we find that
$\tau \simeq 0.2$ for the fireball model ($p=1$) and $\tau \simeq 0.008$ for the AC model
($p=1/3$). A comparison of these estimates with numerical results\citesp{Koers:2007ww} shows that the estimate on $\tau$
is fairly accurate for the AC model over a large range of parameters. For the fireball model, however,
this procedure tends to overestimate the optical depth. The reason for this is that, for a large range of model
parameters, the flow becomes optically thin shortly after pion production becomes possible. This prevents
further acceleration of the flow. Hence the acceleration profile $\Gamma \propto r$ does not hold up to
large radii and the above estimate is not very accurate.
Numerical results indicate that a typical value
for the fireball model is $\tau^{\rm FB} \simeq 0.05$, while for the AC model 
 $\tau^{\rm AC} \simeq 0.01$. Qualitatively, this difference can directly be understood from the dynamics:
 in the AC model,  pion production is only possible at radii an order of magnitude larger than the $np$
 decoupling radius. This implies that the number density of target protons has decreased significantly since
 decoupling,  leading to a small interaction probability. For the fireball model, pion production occurs closer to the decoupling
 radius, where the dilution of target protons is not so strong.
 
\subsection{Neutrino emission}
The neutrino fluence from a single GRB source at proper distance $D_p$ can be expressed as
\mbox{$\Phi_{\nu} \simeq 1.5 N_{n} \tau / 4 \pi D_p^2$}, 
where $N_{n}$ denotes the isotropic-equivalent number of neutrons in the flow, $\tau$ is the $np$ interaction probability,
and we have taken the average number of neutrinos
(adding flavors and antiparticles) per $np$ scattering equal to 1.5.\citesp{Koers:2007ww} Using
\mbox{$N_n \simeq \xi_0 / (1 + \xi_0) \times E/ (\eta m_n c^2)$}, where $\xi_0$ is the initial
neutron-to-proton flux  ratio (cf. eq. \eqref{eq:defxi}), $E$ is the total isotropic-equivalent burst energy,
$\eta$ is the baryon loading parameter (cf. eq. \eqref{eq:defeta}), and $m_n$ the neutron mass,
we find the following neutrino fluences for a burst at redshift $z=1$ for the fireball model and
the AC model, respectively:
\beqs
\label{eq:FBflux}
\Phi_\nu^{\rm FB} \simeq 10^{-4} \,  {\rm cm}^{-2} \lb \frac{\tau}{0.05} \rb \lb \frac{2 \xi_0}{1 + \xi_0} \rb
\lb \frac{E}{10^{53} {\rm erg}} \rb \lb  \frac{\eta}{10^3}\rb^{-1}  \, ; \\
\label{eq:ACflux}
\Phi_\nu^{\rm AC} \simeq 2 \times 10^{-5} \,  {\rm cm}^{-2} \lb \frac{\tau}{0.01} \rb \lb \frac{2 \xi_0}{1 + \xi_0} \rb
\lb \frac{E}{10^{53} {\rm erg}} \rb \lb  \frac{\eta}{10^3}\rb^{-1}  \, .
\eeqs
Using the fact that pions are created near threshold, and assuming a roughly isotropic distribution in the
center-of-mass frame, one finds that the typical observed neutrino energy is $\sim$50 GeV for the fireball model and $\sim$70 GeV for the AC model.\citesp{Koers:2007ww} The typical energy for the AC model is slightly higher because charged pions will be accelerated
by the plasma before decay.
 
For the fireball model, the flux estimate \eqref{eq:FBflux} is roughly an order  of magnitude below
previous estimates\citesp{Bahcall:2000sa}. This difference can be attributed to a more accurate treatment of the $np$
interaction (in Ref.\citesp{Bahcall:2000sa} it is assumed that  $\tau \simeq 1$).
For the AC model, the interaction probability is smaller by another factor $\sim$5. This difference results
from the more gradual acceleration of the flow and is thus directly linked to its nature. 
Unfortunately, the detection prospects with the upcoming
km$^3$ neutrino detectors such as IceCube are very poor due to
the relatively low neutrino energy: for reference values of the parameters we expect less than 1 event per year 
for a combined, diffuse flux of 1000 GRBers per year for either model. This GRB rate is rather optimistic if one takes into account that
$np$ decoupling only occurs for sufficiently pure (high-$\eta$) GRBers. We thus conclude that realistic detection 
prospects for the neutrino flux studied here requires a detector with larger effective area at sub-100 GeV
energies than the upcoming km$^3$ detectors.

\section{Discussion}
Neutrino emission offers a promising way to further our understanding of gamma-ray bursters. Neutrinos
carry information that is complementary to electromagnetic emission because they can escape from optically
thick regions and because they predominantly trace the hadronic component of GRB sources. This offers
a unique way to constrain the nature of the relativistic outflow associated with GRBs. However,
due to their feeble interactions in detectors at Earth, it remains a challenging task to identify concrete realizations of this
potential. 

Here we have discussed neutrino production in inelastic neutron-proton collisions that occur when
neutrons have decoupled from the outflow associated with GRBs. We have estimated the characteristic neutrino flux within
the widely used fireball model and the more recently introduced AC model.
The characteristic neutrino fluxes and energies are distinctively different
for the two models, directly reflecting the dynamics and hence the nature of the flow.
Unfortunately, the relatively low neutrino energy precludes any realistic detection prospects with  the upcoming km$^3$ detectors such as IceCube.

Apart from neutrino production through charged pion decay, one also expects the production of  gamma rays
through the decay of neutral pions produced in $np$ interactions. The plasma is optically thick to these gamma rays,
and hence they cannot directly leave the plasma. In fact, the energy that is injected in the flow through this
mechanism is reprocessed (through synchrotron radiation, pair production, and Inverse Compton scattering)
and emitted in a different energy band. The typical energy of this reprocessed emission is $\sim$10 GeV for
the fireball model and $\sim$100 keV for the AC model, and the expected fluence is
detectable up to large redshifts with the GLAST satellite.\citesp{Koers:2007ww} Detection of this emission would favor the
fireball model, and constrain the baryon loading of the flow.

\section*{Acknowledgments}
It is a pleasure to thank Dimitrios Giannios for a very enjoyable collaboration and for useful comments
on the present manuscript,  and Asaf Pe'er and Ralph Wijers for many valuable discussions on GRB physics.
H.K. is supported by Belgian Science Policy under IUAP VI/11 and by  IISN.

\end{document}